\documentstyle{article}
\begin{document}
\renewcommand{\baselinestretch} {1.5}
\large
\hyphenation{anti-ferro-mag-netic}

\begin{flushright}
{\large{\bf YAMANASHI-94-01\\ May 1994}}
\end{flushright}
\vskip 1.0in
\begin{center}
{\large{\bf  Re-Structuring Method for the Negative Sign Problem\\
in Quantum Spin Systems\\}}
\vskip 1.0in
   Tomo Munehisa and Yasuko Munehisa\\
\vskip 0.5in
   Faculty of Engineering, Yamanashi University\\
   Kofu, Yamanashi, 400 Japan\\
\vskip 1.0in
{\bf Abstract}
\end{center}

We present detailed discussions on a new approach
we proposed in a previous paper to numerically study quantum spin
systems. This method, which we will call
re-structuring method hereafter, is based on rearrangement
of intermediate states in the path integral formulation.
We observed our approach brings remarkable
improvement in the negative sign problem when applied to
one-dimensional quantum spin $1/2$ system with next-to-nearest
neighbor interactions.
In this paper we
add some descriptions on our method and show results
from analyses by the exact diagonalization and by the
transfer matrix method of the system on a small chain.
These results also indicate that our method works quite
effectively.

\eject
\noindent {\bf Section 1  Introduction}

Recent development in experiments on condensed matter has
shown us very interesting systems where strong quantum
effects should be realized.
One of them is the quantum spin system in low dimensions.
Much theoretical work has been done to obtain quantitative as well as
qualitative properties of these quantum effects \cite{rdevel}.

One powerful tool to numerically investigate quantum spin systems
is Monte Carlo approach using the Suzuki-Trotter formula \cite{st}.
Study through this method has brought us very intriguing
results on the ferromagnetic system.
It is widely known, however, if one applies this method to
frustrated systems one often encounters the so-called negative
sign (NS) problem \cite{ns}.
The NS problem, which becomes more serious on larger lattices,
makes it very difficult to get
statistically meaningful results in numerical calculations.

In a previous paper \cite{mune} we showed that a new approach
(re-structuring (RS) method) is useful to obtain Monte Carlo
results for one-dimensional quantum spin $1/2$ system
with next-to-nearest neighbor interactions, which suffers from the
NS problem in the antiferromagnetic case.
Essential point of RS method is to choose a set of states for the
path integral which is appropriate to numerical calculation.
Conventional choice of the set, whose states consist of eigenstates of
$z$-component of the Pauli matrix on each lattice site, is the simplest
one but the NS problem turns out to be serious with this set.
Since any other choice is possible as long as the set is complete,
we employ a set made of eigenstates of local Hamiltonian for
every two neighboring sites.
Although a complete solution for the NS problem could not be obtained,
our results, which show much improvement in numerical calculations,
inspires us with confidence that quantum Monte Carlo method
effectively works
if a set of states for the path integral is appropriately chosen.

In this paper we add descriptions on our method in some detail.
We then analyse the same system by the exact diagonalization and
by the transfer matrix method on a small chain.
We will see these analyses are also helpful to confirm superiority
of RS method to the conventional one.
In section 2 the model and the conventional approach,
which is necessary in our analyses for comparison, are scanned.
Section 3 is for detailed descriptions of RS formulation.
Results from our analyses are given in section 4 and
final section will be devoted to summary and discussions.

\vskip 0.5in
\noindent {\bf Section 2  Model and Conventional approach}

The system we study is
the quantum spin $1/2$ system with next-to-nearest neighbor
interactions on a one-dimensional chain, the simplest one among
those suffering from serious NS problem.
The Hamiltonian of this system is

\begin{eqnarray}
    \hat{H}= {1 \over 2} \sum_{i=1}^N
     (\vec{\sigma}_i\vec{\sigma}_{i+1} +
      \vec{\sigma}_i\vec{\sigma}_{i+2}),
\label{eq:Horig}
\end{eqnarray}
where $N$ is number of sites on the chain
and $\vec{\sigma}_{N+i} \equiv \vec{\sigma}_i$
(periodic boundary condition).
The partition function $Z$ is given by $Z=tr(e^{-\beta \hat{H}})$
with inverse temperature $\beta$.

Let us describe the conventional approach. State on each site
is represented by $z$-component of the spin, namely up and down,
or $+$ and $-$.
In this representation states of the system are given by
  $$    \mid \alpha > = \mid s_1,s_2,s_3,...,s_N > , $$
where $s_i=+$ or $-$. The identity operator is then
  $$  \hat{1} = \sum_{\lbrace s_i \rbrace} \mid s_1,s_2,...,s_N >
      < s_1,s_2,...,s_N \mid . $$
To use this identity operator in the Suzuki-Trotter formula,
we divide the Hamiltonian into four parts\footnote{For
techincal reasons number of sites
in this case should be restricted to be multiple of four.}
as schematically shown in Fig. 1(a). Thus we come to an expression
\begin{eqnarray}
   Z = \lim_{n \rightarrow \infty}
     tr \lbrace (e^{-\beta \hat{H}_1/n }
     e^{-\beta \hat{H}_2/n }
     e^{-\beta \hat{H}_3/n }
     e^{-\beta \hat{H}_4/n } ) ^n \rbrace , \label{eq:Zc}
\end{eqnarray}
where
\begin{eqnarray*}
     \hat{H}_1 =  {1 \over 2} \sum_{i=1}^{N/2}
       \vec{\sigma}_{2i-1}\vec{\sigma}_{2i}, \\
     \hat{H}_2 =  {1 \over 2} \sum_{i=1}^{N/2}
       \vec{\sigma}_{2i}\vec{\sigma}_{2i+1}, \\
     \hat{H}_3 =  {1 \over 2} \sum_{i=1}^{N/4}
       (\vec{\sigma}_{4i-3}\vec{\sigma}_{4i-1} +
        \vec{\sigma}_{4i-2}\vec{\sigma}_{4i}), \\
     \hat{H}_4 =  {1 \over 2} \sum_{i=1}^{N/4}
       (\vec{\sigma}_{4i-1}\vec{\sigma}_{4i+1} +
        \vec{\sigma}_{4i  }\vec{\sigma}_{4i+2}). \\
\end{eqnarray*}
With above complete set and partial Hamiltonians we obtain
the partition function $Z_C^{(n)}$ used in Monte Carlo
calculations,
\begin{eqnarray}
    Z_C^{(n)} = \sum_{\lbrace \alpha_j, \alpha'_j, \alpha''_j,
      \alpha'''_j \rbrace }
   w(\lbrace \alpha_j, \alpha'_j, \alpha''_j,
      \alpha'''_j \rbrace ),   \label{eq:Zcn}
\end{eqnarray}
where
\begin{eqnarray*}
    w(\lbrace \alpha_j, \alpha'_j, \alpha''_j,
      \alpha'''_j \rbrace )  = \prod_{j=1}^n
      <\alpha_j \mid e^{-\beta\hat{H}_1/n} \mid \alpha'_j>
      <\alpha'_j \mid e^{-\beta\hat{H}_2/n} \mid \alpha''_j> \\
    \times
      <\alpha''_j \mid e^{-\beta\hat{H}_3/n} \mid \alpha'''_j>
      <\alpha'''_j \mid e^{-\beta \hat{H}_4/n} \mid \alpha_{j+1}>,
\end{eqnarray*}
suffix $j$ numbering sites along the Trotter axis
and $\alpha_{n+1} \equiv \alpha_1$.

In this system
   $w(\lbrace \alpha_j, \alpha'_j, \alpha''_j,
      \alpha'''_j \rbrace ) $,
total product
of expectation values over one configuration, can be negative.
Appearance of this negative weight brings the NS problem.
In order to numerically calculate some physical quantity
$\langle A \rangle $ one then should subtract
contributions of negatively signed configurations, $A_{-}$,
from those of positively signed ones, $A_{+}$. Namely,
\begin{eqnarray}
   \langle A \rangle ={{A_{+}-A_{-}} \over {Z_{+}-Z_{-}}},
\label{eq:Av}
\end{eqnarray}
where $Z_{+}(Z_{-})$ is number of configurations with positive
(negative) weight. The result would suffer from serious
cancellation when $Z_{-} \simeq Z_{+}$.

\vskip 0.5in
\noindent {\bf Section 3  RS Method}

Formulations of RS method is outlined in our previous paper
\cite{mune}. Here we briefly repeat them and add detailed information
on its effective Hamiltonian.

We start from rewriting $(\ref{eq:Horig})$ in the following form
\begin{eqnarray*}
     \hat{H}= {1 \over 2} \sum_{i=1}^{N/2}
      (\vec{\sigma}_{a,i}\vec{\sigma}_{a,i+1}  +
       \vec{\sigma}_{b,i}\vec{\sigma}_{b,i+1}  +
       \vec{\sigma}_{a,i}\vec{\sigma}_{b,i}    +
       \vec{\sigma}_{b,i}\vec{\sigma}_{a,i+1}),
\end{eqnarray*}
where we denote odd and even sites with suffix $a$ and $b$,
respectively,
   $$\vec{\sigma}_{a,i} \equiv \vec{\sigma}_{2i-1},\ \
     \vec{\sigma}_{b,i} \equiv \vec{\sigma}_{2i}.$$
We employ the complete set for which the operator products
$\vec{\sigma}_{a,i} \vec{\sigma}_{b,i} $ are diagonalized.
Explicitly, we use
\begin{eqnarray}
   \mid \alpha > = \mid S_1,S_2,...,S_{N/2} >,
\label{eq:RSst}
\end{eqnarray}
where $S_i=1_i$, $\oplus_i$, $\ominus_i$ or $-1_i$ with
   $$ \mid 1_i > =\mid +_{a,i} , +_{b,i} > ,$$
   $$ \mid \oplus_i > = {1\over \sqrt{2} }
     (\mid +_{a,i} , -_{b,i} >+ (\mid -_{a,i} , +_{b,i} >),$$
   $$ \mid \ominus_i > ={1\over\sqrt{2} }
     (\mid +_{a,i} , -_{b,i} >- (\mid -_{a,i} , +_{b,i} >),$$
   $$ \mid -1_i > = \mid -_{a,i} , -_{b,i} >  . $$
Hamiltonian $(\ref{eq:Horig})$ is accordingly devided into
``odd'' and ``even'' parts
\begin{eqnarray}
   \hat{H}=\hat{H_o}+\hat{H_e},
\label{eq:Hoe}
\end{eqnarray}
where, as shown in Fig. 1(b),
\begin{eqnarray}
   \hat{H_o} = {1 \over 2} \sum_{i=1}^{N/4} \hat{h}_{oi}
\end{eqnarray}
\begin{eqnarray*}
   \hat{h}_{oi}=
     (\vec{\sigma}_{a,2i-1} \vec\sigma_{a,2i}
     + \vec\sigma_{b,2i-1} \vec\sigma_{b,2i}
     + \vec\sigma_{b,2i-1} \vec\sigma_{a,2i} \\
   + {1 \over 2} \vec\sigma_{a,2i-1} \vec\sigma_{b,2i-1}
     + {1 \over 2} \vec\sigma_{a,2i} \vec\sigma_{b,2i} ),
\end{eqnarray*}
\begin{eqnarray}
   \hat{H_e} = {1 \over 2} \sum_{i=1}^{N/4} \hat{h}_{ei} ,
\end{eqnarray}
\begin{eqnarray*}
   \hat{h}_{ei} =
     (\vec{\sigma}_{a,2i} \vec\sigma_{a,2i+1}
     + \vec\sigma_{b,2i} \vec\sigma_{b,2i+1}
     + \vec\sigma_{b,2i} \vec\sigma_{a,2i+1} \\
   + {1 \over 2} \vec\sigma_{a,2i-1} \vec\sigma_{b,2i-1}
     + {1 \over 2} \vec\sigma_{a,2i} \vec\sigma_{b,2i} ).
\end{eqnarray*}
Expression of the partition function with these partial
Hamiltonians is
\begin{eqnarray}
     Z=\lim_{n \rightarrow \infty} tr \lbrace
      (e^{-\beta \hat{H}_o/n} e^{-\beta \hat{H}_e/n} )^n \rbrace .
\label{eq:Zrs}
\end{eqnarray}
Inserting identity operators made of $\alpha$'s in $(\ref{eq:RSst})$
between the exponents we obtain partition function in RS method,
\begin{eqnarray}
   Z_{RS}^{(n)} = \sum_{ \lbrace \alpha_j,\alpha'_j \rbrace }
     \prod_{j=1}^{n} <\alpha_j \mid e^{-\beta \hat{H}_o/n}
     \mid \alpha'_{j} > <\alpha'_j \mid
     e^{-\beta \hat{H}_e/n} \mid \alpha_{j+1}>.
\label{eq:Zrsn}
\end{eqnarray}
In order to obtain Boltzman weights to be used in
Monte Carlo simulations, we need to construct
$<S'_{2i-1},S'_{2i} \mid exp(-{\beta \over 2n } \hat{h}_{oi})
\mid S_{2i-1},S_{2i}>$ from local matrix elements
$<S'_{2i-1},S'_{2i} \mid \hat{h}_{oi} \mid S_{2i-1},S_{2i}> $
as well as
$<S'_{2i},S'_{2i+1} \mid exp(-{\beta \over 2n } \hat{h}_{ei})
\mid S_{2i},S_{2i+1}> $ from local matrix elements
$<S'_{2i},S'_{2i+1} \mid \hat{h}_{ei} \mid S_{2i},S_{2i+1}>$.
Since these matrix elements are $i$-independent and
$$<S'_{2i-1},S'_{2i} \mid \hat{h}_{oi} \mid S_{2i-1},S_{2i}> =
<S'_{2i},S'_{2i+1} \mid \hat{h}_{ei} \mid S_{2i},S_{2i+1}>, $$
we only need to know values of
$$<S'_1,S'_2 \mid \hat{h}_{o1} \mid S_1,S_2> =
<S'_1,S'_2 \mid \vec{\sigma}_{a,1} \vec{\sigma}_{a,2} \mid S_1,S_2> $$
$$+<S'_1,S'_2 \mid \vec{\sigma}_{b,1} \vec{\sigma}_{b,2} \mid S_1,S_2>
+<S'_1,S'_2 \mid \vec{\sigma}_{b,1} \vec{\sigma}_{a,2} \mid S_1,S_2> $$
\begin{eqnarray}
+{1 \over 2}
<S'_1,S'_2 \mid \vec{\sigma}_{a,1} \vec{\sigma}_{b,1} \mid S_1,S_2>
+{1 \over 2}
<S'_1,S'_2 \mid \vec{\sigma}_{a,2} \vec{\sigma}_{b,2} \mid S_1,S_2> .
\label{eq:mxel}
\end{eqnarray}
It is an easy task to calculate each matrix element in (\ref{eq:mxel}).
For example,
$$<-1_1,1_2 \mid \vec{\sigma}_{a,1} \vec{\sigma}_{a,2}
\mid \oplus_1,\ominus_2> $$
$$=<-_{a,1},-_{b,1} \mid <+_{a,2},+_{b,2} \mid
(2\sigma_{a,1}^+ \sigma_{a,2}^-
+2\sigma_{a,1}^- \sigma_{a,2}^+
+\sigma_{a,1}^z \sigma_{a,2}^z) $$
$${1 \over \sqrt{2}} (\mid +_{a,1},-_{b,1}> + \mid -_{a,1},+_{b,1}> )
{1 \over \sqrt{2}} (\mid +_{a,2},-_{b,2}> - \mid -_{a,2},+_{b,2}> ) $$
$$=<-1_1,1_2 \mid
(\mid 1_1,-1_2> - \mid -1_1,1_2> + \mid \ominus_1, \oplus_2>)
= -1,$$
where
$$\sigma^{\pm}={1 \over 2} (\sigma^x \pm i\sigma^y). $$
Values of
$<S'_1,S'_2 \mid \hat{h}_{o1} \mid S_1,S_2>$ are summarized in Table.

\eject
\noindent {\bf Section 4  Transfer Matrix and Exact Diagonalization}

In this section we present results from analyses
with RS partition function $(10)$ on an $N=8$ chain
by the transfer matrix method \cite{tr} and the exact diagonalization.
Results from conventional partition function $(3)$ are also provided
for comparison.
Purpose of these analyses are twofold.
One is to confirm RS algorithm and previous Monte Carlo results.
Another is to study how systems with $Z_C^{(n)}$ and $Z_{RS}^{(n)}$
depend on Trotter number $n$.

First let us show results for ratio of negatively weighted
configurations to total configurations,
$P=Z_-/(Z_+ + Z_-)$, calculated by the transfer matrix method.
We use `real-space' transfer matrix to realize
the same boundary condition as those in other methods.
In the conventional approach we calculate product of matrix $T_k$,
whose matrix element $(T_k)_{ \alpha' \alpha}$  corresponds
to $<\alpha' \mid exp( -\beta \hat{H}_k /n ) \mid \alpha >$.
Therefore
\begin{eqnarray*}
 Z_C^{(n)} = tr((T_1 T_2 T_3 T_4 )^n).
\end{eqnarray*}
Similarly, with $(T_{o(e)})_{ \alpha' \alpha}=
<\alpha' \mid exp( -\beta \hat{H}_{o(e)} /n ) \mid \alpha >,$
\begin{eqnarray*}
 Z_{RS}^{(n)} = tr((T_e T_o  )^n).
\end{eqnarray*}

In figure 2(a) we plot $P$ from the transfer matrix method together with
previous Monte Carlo results. We see Monte Carlo data and transfer
matrix results for $n=2$ agree very well. Also we observe agreement
between those for larger Trotter numbers is
satisfactory\footnote{Small discrepancy in low temperature region
might suggest that additional improvements in global flips,
which are related to the ergodicity\cite{erg}, are necessary.}.
Our Monte Carlo work is thus nicely supported by an analytical
calculation.

Next we calculate
\begin{eqnarray}
    R={{Z_+ - Z_-} \over {Z_+ + Z_-}}={Z \over Z'},
\label{eq:Rpm}
\end{eqnarray}
which equals to $1-2P$ and is the
ratio of the partition function $Z$ obtained from the
Hamiltonian $(\ref{eq:Horig})$
to the partition function $Z'$ calculated from the modified
Hamiltonian that gives absolute weights \cite{ns}.

For $N=8$ chain it is not difficult to calculate all energy eigen
values $E_i$ ($i=1,...,2^N$) by the exact diagonalization.
We therefore can obtain value of exact partition function at any
temperature by equation
\begin{eqnarray}
  Z =  \sum_{i=1}^{2^N} exp( -\beta E_i).
\label{eq:extZ}
\end{eqnarray}
The same is true for the denominator in $(\ref{eq:Rpm})$;
\begin{eqnarray}
  Z' =  \sum_{i=1}^{2^N} exp( -\beta E'_i),
\label{eq:absZ}
\end{eqnarray}
with energy eigen values $E'_i$ for the modified Hamiltonian.
We refer to the ratio $R$ calculated from $(\ref{eq:extZ})$ and
$(\ref{eq:absZ})$ as $R_{DG}$. It should be noted
that value of $R_{DG}$ depends on method used to
calculate the denominator $Z'$.
We will see RS method gives better ratio compared
to the conventional method.

Before going to full numerical calculations, let us comment on
ground state energy of each Hamiltonian.
As discussed in ref.\cite{ns}, at very low temperature
where contribution from the lowest energy eigen value is dominant,
$$  R \simeq  exp( -\beta (E_0 - E'_0)) $$
because
$  Z_+ -  Z_- \simeq  exp(-\beta E_0)  $ and
$  Z_+ +  Z_- \simeq  exp(-\beta E'_0) $
with $E_0$ $(E'_0)$ denoting energy of ground state for the
Hamiltonian (for the modified Hamiltonian).
Existence of severe NS problem implies $ E_0 > E'_0$.

Our results are in agreement with the above discussion. Values
we obtain by the exact diagonalization are
$   E_0 = -8.25 $ and
$   E'_0 = -10.14 $ $(-11.32)$ for RS (convensional) method.
The fact that the energy difference $E_0 - E'_0$ in RS method is
smaller than that in the conventional method indicates improvement
introduced by RS method.

For the second purpose we also calculate $R$ by the transfer matrix
method with several values of Trotter number $n$.
All numerical results on $R$ are summarized in figure 2(b).
Here we would like to point out two features. One is that,
as is expected,
the $R_{DG}$ is much larger in the RS method than
in the convensional method, especially for low temperatures.
Another feature is rather unexpected one. In RS method it turned
out that $R$'s for finite Trotter numbers, which should coincide with
$R_{DG}$ in $n \rightarrow \infty$ limit, are always larger than
$R_{DG}$. In the conventional method, on the contrary, all $R$
values calculated by the transfer matrix method lie below $R_{DG}$.
This encourages us to promote RS method in Suzuki-Trotter formula
because it becomes possible to obtain statistically meaningful Monte
Carlo results even for small values of $n$.

Finally we present results for the energy of the system. In figure 3,
exact values obtained by the exact diagonalization,
$$  E = \sum_{i=1}^{2^N} E_i \cdot e^{-\beta E_i}
      / \sum_{i=1}^{2^N} e^{-\beta E_i}, $$
and other values calculated by the transfer matrix method are plotted.
It is clear that conventional method fails to give reliable values
when Trotter number is small. Even when $n=8$, difference from
exact one becomes large as temperature $T$ decreases.
In the RS method, on the contrary, we get satisfactorily stable
results even for $n=2$ and all data seem to nicely converge to
the exact ones when $n$ is enlarged.

\vskip 0.5in
\noindent {\bf Section 5  Discussions}

In this paper we provide additional explanations of RS approach,
which we proposed in a previous paper \cite{mune} in order to make
meaningful Monte Carlo study of quantum spin systems
suffering from the NS problem.
 From technical point of view main task is to analytically
calculate matrix elements for effective Hamiltonian in
$(\ref{eq:Zrsn})$.
Descriptions on it are presented in section 3.
We also present results from analyses of
quantum spin $1/2$ system with next-to-nearest neighbor
antiferromagnetic interactions on a small chain, applying
RS approach in the transfer matrix method and the exact diagonalization.
In section 4 we see all results obtained through these
analyses indicate that RS method is effectively works in this case.

Let us comment on possible applications of RS method to other
quantum spin systems.
Qualitatively the RS method will be effective for any
frustrated systems suffering from the NS problem.
But its power to bring how much quantitative improvement
in numerical calculations does strongly depend on system's
interactions and spacial constructions of clusters.

One interesting application would be to investigate
quantum spin $1/2$ XY chain
with next-to-nearest neighbor interactions.
For qualitaitve discussion we calculate energy eigen values
of the system on $N=8$ chain by the exact diagonalization.
The lowest enegy is $E_0=-5.78$ for the XY Hamiltonian
while $E'_0=-7.14(-9.44)$
for the modified Hamiltonian in RS (conventional) method.
These results suggest RS method works more powerfully
on the XY chain in comparison with the Heisenberg chain,
which should be one example of the above statement.

Another intriguing application was made
to the $\Delta$ chain
by Nakamura and Saika.
They reported that RS method is quite useful for this system
\cite{nakamura}.

We would like to point out a very lucky case where
the NS problem is completely solved, although it might be less
interesting from physical point of view.
This is one-dimensional antiferromagnetic quantum spin $1/2$ system
with next-to-next-to-nearest neighbor interactions in addition to
next-to-nearest neighbor one, where all
coulpings have the same strength\footnote{Number of sites of this system
also should be multiple of four for technical reasons.}.
This system sufferes from severe NS problem in the conventional
approach.
When one chooses eigen states of the nearest interaction for the nearest
two spins, matrix elements of the effective Hamiltonian
become quite simple so that negatively weighted configurations
do not appear at all.

In two-dimensional cases systems on triangular lattice will be
most important in application.
If we employ eigen states for cluster of two
neighboring spins as we did in one-dimensional case
it is not unique which two should form a pair because
more than one neighboring sites exist on the lattice.
Symmetry of the lattice inspires us clustering of three
spins is appropriate, for which one has to do with more
complicated effective matrix elements.
At present it is an open question which way gives better
results.

In any case a good choice of complete set which
improves the NS problem will help us to gain deep insight into
the ground state of the system under investigation.

\eject
\noindent
{\bf Acknowledgements}

We would like to express sincere thanks to Drs. T. Nakamura and N. Hatano,
who made quick responce to our previous work and sent valuable comments
with their theses that turned out very useful for our study.
Also we are deeply grateful to Profs. S. Miyashita and M. Imada for
nice review of quantum Monte Carlo study in Kotai-Butsuri (Japanese).

\eject

\eject
\noindent {\bf Table Caption}

Values of matrix elements
$<S'_1,S'_2 \mid \hat{h}_{o1} \mid S_1,S_2>$
described in section 3.
Rows and colomns denote states $<S'_1,S'_2 \mid $ and
$\mid S_1,S_2>$, respectively.
All other matrix elements are zero.

\vskip 1.0in
\noindent {\bf Figure Captions}

\noindent {\bf Figure 1} \\
(a) Schema to show how to divide the Hamiltonian
$(\ref{eq:Horig})$ into four parts
on an eight-site $(N=8)$ chain with periodic boundary condition.
Sites are numbered from one to eight.
Filled circles, squares, diamonds and triangles on
links denote interactions to belong partial Hamiltonians
$\hat{H}_1$, $\hat{H}_2$, $\hat{H}_3$ and
$\hat{H}_4$ in $(\ref{eq:Zc})$, respectively. \\
(b) Schema to show how to re-divide the same Hamiltonian
$(\ref{eq:Horig})$ into $\hat{H}_o$ and $\hat{H}_e$ in
$(\ref{eq:Zrs})$.
Sites are renumbered from $a1$ to $b4$.
Open diamonds (triangles)
denote interactions to fully belong to partial Hamiltonian
$\hat{H}_o$ ($\hat{H}_e$). Interactions on links with open
circles should be equally shared by $\hat{H}_o$ and $\hat{H}_e$.

\noindent {\bf Figure 2} \\
(a) Ratio of negative weight calculated on an $N=8$ chain
by Monte Carlo simulations (MC) and by the transfer matrix method (TM),
using the RS Hamiltonian (RS) or the conventional one (conv.),
with several values of Trotter number $n$. \\
(b) Ratio of partition functions on an $N=8$ chain.
The exact diagonalization (DG) is employed in addition to the
transfer matrix method.

\noindent {\bf Figure 3} \\
Energy per site calculated on an $N=8$ chain by means of the exact
diagonalization (DG)
and the transfer matrix method (TM). The transfer matrix method is applied
to both Hamiltonians (RS, conv.) with Trotter number $n=$2, 4 and 8.

\eject
\noindent {\bf Table}
\vskip 0.3in
\begin{tabular}{|c|c|}  \hline
  $       $ & 1, 1 \\ \hline
  1, 1      & 4  \\ \hline
\end{tabular}

\vskip 0.2in

\begin{tabular}{|c|c|c|c|c|}  \hline
  $       $ & $\oplus$,1 & $\ominus$,1 & 1,$\oplus$ & 1,$\ominus$ \\ \hline
  $\oplus$,1     & 1 &  1 &  3 &  1 \\ \hline
  $\ominus$,1    & 1 & -1 & -1 &  1 \\ \hline
  1,$\oplus$     & 3 & -1 &  1 & -1 \\ \hline
  1,$\ominus$    & 1 &  1 & -1 & -1 \\ \hline
\end{tabular}

\vskip 0.2in

\begin{tabular}{|c|c|c|c|c|c|c|}  \hline
  $       $ & -1, 1 &  1, -1  &
$\oplus$, $\oplus$ &  $\ominus$, $\oplus$ &
$\oplus$, $\ominus$ & $\ominus$, $\ominus$  \\ \hline
  -1, 1               & -2 & 0  & 3 & 1  & 1  & -1 \\ \hline
   1,-1               & 0  & -2 & 3 & -1 & -1 & -1 \\ \hline
 $\oplus$, $\oplus$   & 3  & 3  & 1 & 0  & 0  & 1  \\ \hline
  $\ominus$, $\oplus$ &  1 & -1 & 0 & -1 & 1  & 0  \\ \hline
 $\oplus$, $\ominus$  &  1 & -1 & 0 & 1  & -1 & 0  \\ \hline
 $\ominus$, $\ominus$  & -1 & -1 & 1 & 0  & 0  & -3 \\ \hline
\end{tabular}
\vskip 0.2in
\begin{tabular}{|c|c|c|c|c|}  \hline
  $ $ & $\oplus$,-1 & $\ominus$,-1 & -1,$\oplus$ & -1,$\ominus$ \\ \hline
  $\oplus$,-1     & 1  & -1 & 3 & -1 \\ \hline
  $\ominus$,-1    & -1 & -1 & 1 & 1  \\ \hline
  -1,$\oplus$     & 3  &  1 & 1 & 1  \\ \hline
  -1,$\ominus$    & -1 &  1 & 1 & -1 \\ \hline
\end{tabular}
\vskip 0.2in
\begin{tabular}{|c|c|}  \hline
  $       $ & -1, -1  \\ \hline
  -1, -1     & 4      \\ \hline
\end{tabular}
\end{document}